\documentstyle[aps,prd]{revtex}

\def\be{\begin{equation}}
\def\ee{\end{equation}}
\def\beq{\begin{eqnarray}}
\def\eeq{\end{eqnarray}} 
\def\s{\sigma}

\def\G{\Gamma} 
 
\def\an{analytic}
\def\ac{\an{} continuation} 
\def\hsr{hypergeometric series representations} 
 
\def\ndim{NDIM} 
 
\def\half{\frac{1}{2}} 
\def\lci{light-cone gauge integrals}

\begin{document}

\draft
\title{Light-cone gauge integrals: Prescriptionlessness at two loops}
\author{Alfredo T. Suzuki\footnote{E-mail:suzuki@ift.unesp.br}, Alexandre G. M. 
Schmidt\footnote{E-mail:schmidt@ift.unesp.br}}
\address{Universidade Estadual Paulista --- Instituto de F\'{\i}sica Te\'orica \\ 
R.Pamplona, 145 S\~ao Paulo - SP CEP 01405-900, Brazil}
\date{\today}\maketitle

\begin{abstract}
The only calculations performed beyond one-loop level in the
light-cone gauge make use of the Mandelstam-Leibbrandt (ML)
prescription in order to circumvent the notorious gauge dependent
poles. Recently we have shown that in the context of negative
dimensional integration method (\ndim{}) such prescription can be
altogether abandoned, at least in one-loop order calculations.  We
extend our approach, now studying two-loop integrals pertaining to
two-point functions. While previous works on the subject present only
divergent parts for the integrals, we show that our prescriptionless
method gives the same results for them, besides finite parts for
arbitrary exponents of propagators.

\end{abstract} 

\vspace{.5cm}



\section{Introduction}

Perharps the trickiest gauge in what someone has termed the
``gauge-market'' \cite{leib-rmp} is the light-cone gauge. The problem
of spurious poles arising in such a choice is well-known and the
insights gained throughout the years, concerning the problems related
to naive implementation of dimensional regularization or Cauchy
principal value to treat them, are registered in the pertinent
literature as well as the ultimate remedy\cite{mandel,pimentel} for
the pathologies, that so tenaciously has defied for many years our
correct understanding of them: causal prescriptions. 

Prescriptions may be of good help for a time and as long as they can
solve immediate problems. If, on the other hand we could somehow avoid
them and at the same time understand why we can dispense with them
would be no doubt a better picture to envisage. Moreover, using such
prescriptions turn the calculations very laborious --- partial
fractioning and integration over components are, in fact, part of the
ML prescription. On the other hand, negative dimensional integration
method (\ndim{}) can dismiss the referred tricks and more: integrals
can be evaluated for arbitrary exponents of propagators and
dimension. Always preserving causality and gauge invariance.

Our aim is to iron out \ndim{}, that is, to perform a two-loop test to
our prescriptionless method for \lci{}. The aftermath of this is the
establishment of a growing confidence in the novel methodology, which
thoroughly dispense with prescriptions as well as partial fractioning
of two-loop light-cone singularities that arise at this level of
perturbative calculations.  We give results in terms of hypergeometric
series for arbitrary exponents of propagators and dimension, which
enable us to select particular values for them and especially those
that are relevant to physical amplitudes, with not only the pole
pieces of integrals, but their finite parts as well.

\section{Light-cone Gauge at Two-loops: a toy integral in \ndim{}}

The principle of gauge invariance is foundational in modern quantum
field theory. It assures that gauge fields are invariant under gauge
transformations. In fact, we can go so far as to say that the gauge
principle is the key to our present understanding of elementary
particles and their interactions. A related but very often
misunderstated concept is the question of gauge independent
quantities. This has to do with the fact that relevant physical
quantities of interest in any phenomenon must be {\em gauge
independent}, i.e., whatever the gauge choice we make use of, the
resulting measurable quantity must be independent of the choice
made. In other words, calculations of physical quantities performed in
any gauge must give the same result. 

In this trail, we can say that the gauge choice we make is more or
less a matter of taste and preference, and of practical reasons
related to the amount of hard work we have to make in order to achieve
the end results. The lure of the light-cone gauge has its reason,
since it is a physical gauge, meaning that ghost fields decouple from
matter and gauge fields, unitarity is preserved without auxiliary
scalar fields with the ``wrong'' fermionic statistics, and the gauge
propagator has a deceivingly ``simple'' structure. Yet, attached to
the same coin, the other side of it has subtleties peculiar to this
gauge.

Until Mandelstam and Leibbrandt introduced their prescriptions ---
which are in fact equivalent --- loop integrals in the physical
light-cone gauge were not well understood since calculations of
Feynman diagrams in such a choice produced ill-defined integrals
proportional to $\Gamma (0)$ and integrals with pathological results
such as double poles in single loop diagrams. Later on, all these
pathologies were understood to be related to violation of causality
associated to naive use of dimensional regularization or the use of
principal value prescription to deal with light-cone singularities.

The physical light-cone gauge is defined through an external,
constant, light-like four-vector $n^\mu,\:\:\:n^2=0$, but it is now
well-understood that this single vector is not sufficient to span the
whole four-dimensional space-time, which in other words means that the
gauge freedom is not totally removed. Residual gauge freedom remains
to be removed by the introduction of the dual four-vector
$n^{*\mu},\:\:\: n^{*2}=0$\cite{tetrad}. This is done by {\em hand} in the
ML-prescription whereas in our approach this is done naturally as a
consequence of the general structure of the light-cone integral,
defined over four-dimensional Minkowski space-time.

Our negative-dimensional approach can avoid the use of
prescriptions\cite{without} and provide physically acceptable results,
i.e., causality preserving ones. The calculation we will present is
the very first test beyond one-loop order without invoking the
ML-prescription. We demonstrate that integration over components and
partial fractioning tricks can be completely abandoned as well as
parametric integrals. The important point to note\cite{without} is
that the dual light-like four-vector $n^*_\mu$ is necessary in order
to span the whole four-dimensional space\cite{tetrad,leib-cjp}, when
defining the gauge proper. Of course all these are, in the course of
calculations dimensionally regularized into a $D$-dimensional
space-time.

So far, we have tested our \ndim{} for integrals pertaining to
one-loop class. Now we apply such technique to some massless two-loop
integrals. Let us consider an integral studied by Leibbrandt and
Nyeo\cite{leib-nyeo}. The reason for this is two-fold: Firstly, they
did not calculate it explicitly, but gave only the pole structure of
it, showing that it does not contain any pathological features; and
secondly because in our approach, a whole class of integrals is
calculated simultaneously, their integral being a particular case of
ours.

Let us define the following \lci{},

\be 
C_3 = \int d^D\! q\, d^D\! k\, \frac{k^2}{q^2 (q-k)^2 (k-p)^2 (k\cdot
n) (q\cdot n)},
\ee
where in their calculation the ML-prescription 
\be 
\frac{1}{k\cdot n} = \lim_{\epsilon\to 0} \frac{k\cdot n^*}{(k\cdot
n)(k\cdot n^*) +i\epsilon}, 
\ee
must be understood. On the other hand, in the \ndim{} context the key
point is to introduce the dual vector $n_\mu^*$ in order to span the
complete space\cite{tetrad,without,leib-cjp}. If we do not consider it,
our result will violate causality, giving the Cauchy principal value
of the integral in question, as we concluded in \cite{probing}.

Our aim to is perform,

\be \label{N}
{\cal N} = \int d^D\! q\, d^D\! k_1\, (k_1^2)^i (q^2)^j (q-k_1)^{2k}
(k_1-p)^{2l} (k_1\cdot n)^m (q\cdot n)^s (k_1\cdot n^*)^r. 
\ee

{\em The attentive reader will notice here that we have only put the
factor $(k_1\cdot n^*)^r$ and the other possible factor $(q\cdot
n^*)^t$ is conspicuosly absent. The latter could be taught as
necessary on the grounds of equal footing ascribed to integrals in
$k_1$ and $q$. However, the \ndim{} technique requires only the former
one. The reason for this is related to the fact that the integrals in
$k_1$ and in $q$ must generate factors of the form $p\cdot n^*$ of the
external momentum to ensure the complete spanning of the physical
four-dimensional space-time \cite{tetrad,leib-cjp}. From (\ref{N}) we
can readily see that only the $k_1$ integration contains the
propagator $(k_1-p)^{2l}$ that can generate the needed $p\cdot n^*$ in
the external momentum, whereas the $q$ integration is unable to do so,
and the inclusion of $q\cdot n^*$ is completely unnecessary. So here
again we can see the power of \ndim{}.  Although we treat each
integration on the same footing, the propagators must be treated in a
way to ensure the correct spanning of the physical four-dimensional
space-time in the end result.}

We will carry out this integral and then present results for special
cases, including Leibbrandt and Nyeo's $C_3$, where $i=1$, $r=0$ and
the other exponents equal to minus one.  Observe that the integral
must be considered as a function of external momentum, exponents of
propagators and dimension,

\be 
{\cal N} = {\cal N}(i,j,k,l,m,r,s;P,D), 
\ee
where $P$ represents $(p^2, p^+, p^-, \half (n\cdot n^*))$, and we
adopt the usual notation for the light-cone gauge\cite{leib-rmp}.

Our starting point is the generating functional for our
negative-dimensional integrals,

\be 
G_N = \int d^D\! q\, d^D\! k\, \exp{\left[-\alpha k^2 -\beta q^2
-\gamma (q-k)^2 -\theta (k-p)^2 -\phi (k\cdot n) -\omega (q\cdot n)
-\eta (k\cdot n^*)\right]},
\ee
which after a little bit of algebra and integration over momenta yields,

\beq 
G_N &=& \left(\frac{\pi^2}{\lambda}\right)^{D/2} \!\!\!\exp \left \{
\frac{1}{\lambda} \left[ -g_1p^2 -g_2(p\cdot n) -g_3 (p\cdot n^*) +g_4
\left(\half n\cdot n^*\right)\right] \right\} ,
\eeq
where
$$g_1 = (\alpha\beta + \alpha\gamma +\beta\gamma)\theta, \quad g_2 =
(\beta\phi+ \gamma\omega +\gamma\phi) \theta,\quad g_3 =
(\beta+\gamma)\eta\theta, \quad g_4 = \eta\frac{g_2}{\theta},$$ and $
\lambda = \alpha\beta+ \alpha\gamma + \beta\gamma + \beta\theta +
\gamma\theta.$

Taylor expanding directly the exponentials we obtain,

\be 
{\cal N} = (-\pi)^D i!j!k!l!m!r!s!\G(1-\s-D/2) \sum_{{\rm
all}=0}^\infty \frac{ \delta}{ X_1!\dots
X_8!Y_1!Y_2!Y_3!}\frac{(p^2)^{X_{123}}(p^+)^{X_{456}}(p^-)^{X_{78}}}{Z_1!
\dots Z_5!}\left( -\frac{n \cdot n^*}{2} \right)^{Y_{123}}, 
\ee
where $\s=i+j+k+l+m+r+s+D$, $X_{123}=X_1+X_2+X_3$, etc. and $\delta$
represents the system of constraints $(8\times 16)$ for the
negative-dimensional integral. At the end of the day we have $12,870$
possible solutions for such system. Most of them, $9,142$, have no
solution while $3,728$ present solutions which can be written as
hypergeometric series. Of course several of these will provide the
same series representation, and these solutions we call degenerate.

First of all, we present a result for the referred integral as a double
hypergeometric series,

\beq 
{\cal N}_2^A &=& \pi^D f_2^A P_2^A \sum_{Z_j=0}^\infty
\frac{(\s_n+D/2|Z_{45}) (i+j+k+m+s+D|Z_{45})
(D/2+k|Z_4)}{Z_4!Z_5!(1+i+j+k+\s_n+D|Z_{45})(j+k+s+D|Z_{45})}
\frac{(j+s+D/2|Z_5)}{(1+i+j+k+m+r+s+D|Z_{45})} \nonumber\\ &&\times
(i+j+k+r+D|Z_{45})\left(\frac{p^2n\cdot n^*}{2p^+
p^-}\right)^{Z_{45}},
\eeq
where 
\beq 
f_2^A &=& (-m|-s)(-i-j-k-D/2|-\s_n-D/2)
(j+k+s+D|i-s+r)\frac{(-l|k+l+D/2) (-k|-j-D/2)
}{(1+r|-i-j-k-m-r-s-D)}\\ &&\times
(-m|j+m+s+D/2)(-j|-i-k-m-r-s-D),\nonumber 
\eeq
are the coefficients in terms of Pochhammer symbols, $(x|y)\equiv
(x)_y=\G(x+y)/\G(x)$ and
 
\be 
P_2^A = (p^2)^{\s_n+i+j+k+D} (p^+)^{l+m+s-\s_n}(p^-)^{l+r-\s_n}
\left(\frac{n\cdot n^*}{2} \right)^{\s_n-l},
\ee
is the external kinematical configuration.

Observe that in the above series we must have $|z|<1$, where
$z=p^2n\cdot n^*/2p^+p^-$, in order to be convergent. Now we can
consider the special case ($i=1, j=k=l=m=s=-1, r=0$), studied in
\cite{leib-nyeo},

\beq\label{caso-especial} 
C_3 &=& \pi^D \frac{\G(5-2D) \G(D-1) \G(D/2-1)
\G(2-D/2)\G(D/2-2)}{\G(1-D/2)\G(D-3)} (p^2)^{2D-5}(p^+)^{1-D}
(p^-)^{3-D} \left(\frac{n\cdot n^*}{2} \right) ^{D-3} \nonumber\\
&&\times\sum_{Z_4,Z_5=0}^\infty
\frac{(3D/2-4|Z_{45})}{Z_4!Z_5!(2D-4|Z_{45})}
\frac{(D/2-1|Z_4)(D/2-2|Z_5)(D-1|Z_{45})}{(D-2|Z_{45})}
\left(\frac{p^2 n\cdot n^*}{2p^+p^-} \right)^{Z_{45}} ,
\eeq 
which clearly exhibits a double pole, as stated by Leibbrandt and
Nyeo\cite{leib-nyeo}.

Outside the region $|z|<1$, there is a solution which in principle
could be obtained by \ac{} from the above result, provided the formulae
for such \ac{} were known. However, as far as we are aware of, this is
not the case. \ndim{} on the other hand, gives several \hsr{} for
Feynman loop integrals which are all connected by \ac{}, and \ac{}
formulas can in principle be implied from them. We quote only the
final result valid when $|z|>1$,

\be\label{2som-inv} 
{\cal N}_2^B = \pi^D f_2^B P_2^B \sum_{X_i=0}^\infty \frac{(D/2+k|X_7)
(D/2+j+s|X_8) (-r|X_{78}) }{X_7!X_8! (1+m-r+s|X_{78})
(1-l-r-D/2|X_{78})} \frac{(i+j+k+m+s+D|X_{78})}{(D+j+k+s|X_{78})
}\left(\frac{2p^+p^-}{p^2 n\cdot n^*} \right)^{X_{78}},
\ee
where
\beq 
f_2^B &=& (-m-s|r) (\s+D/2|-l-r-D/2)
(-i-j-k-D/2|-l-r-D/2)(-l|j+l+s+D/2)\nonumber\\ &&\times
(j+k+s+D|-j-s-D/2) (-j|-k-D/2) (-k|k+l+r+D/2),
\eeq
and $ P_2^B = (p^2)^{i+j+k+l+r+D} (p^+)^{m-r+s}$. An important point
to observe here is that, hypergeometric series does not allow negative
integer denominator parameters, for they are not well defined in this
case. Care must therefore be taken because of the factor
$(1+m-r+s|X_{78})$ present in the denominator.

There are also results given as triple hypergeometric series,

\be
{\cal N}_3^A = \pi^D P_1 \Gamma_1 {\cal F}_1\left(\frac{p^2n\cdot
n^*}{2p^+p^-} \right) + \pi^D P_2 \Gamma_2
{\cal F}_2 \left(\frac{p^2n\cdot n^*}{2p^+p^-} \right),\label{3som-inv}
\ee
where 

\beq
{\cal F}_1\left(\frac{p^2n\cdot n^*}{2p^+p^-} \right)\equiv
\sum_{Y_i=0}^\infty \frac{(D/2+l|Y_{123})(-r|Y_{123}) (-m|Y_{12})
(D/2+k|Y_1)(-s|Y_3)(D/2+j+s|Y_2) }{Y_1!Y_2!Y_3! (1-\s +l|Y_{123})
(D+j+k+s|Y_{12}) (1+i+j+k+l+D|Y_{123})} \left(\frac{p^2n\cdot
n^*}{2p^+p^-} \right)^{ Y_{123}},
\eeq
and

\beq
{\cal F}_2\left(\frac{p^2n\cdot n^*}{2p^+p^-}\right )& \equiv &
\sum_{X_i=0}^\infty \frac{(D/2+k|X_1)(-i|X_{12}) (-\s+r|X_{123})
(-i-j-k-l-r-D|X_{123}) (-j-k-D/2|X_3)}{X_1!
X_2!X_3!(1-i-j-k-l-D|X_{123})} \nonumber\\ & \times &
\frac{(j+s+D/2|X_2)}{ (1-\s-i-j-k-D|X_{123})
(j+k+s+D|X_{12})}\left(\frac{p^2n\cdot n^*}{2p^+p^-} \right)^ {
X_{123}}.
\eeq

The coefficient factors of external momenta are defined in the
following table and the Pochhammer symbols are,

\beq 
\G_1 &=&
(-l|\s)(-i-j-k-D/2|i)(j+k+s+D|-k-D/2)(-j|-i-k-l-D)(\s+D/2|-\s+k)\nonumber\\
&\times &(-k|k+l+D/2),
\eeq
and
\beq 
\G_2 &=& (\s+D/2|i+j+k+D/2)(j+k+s+D|-k-D/2)(-m-s|\s-r)(-j|-k-D/2)\nonumber \\
&\times &\frac{(-k|-i-j-l-r-D)(-l|k+l+D/2)}{(1+r|\s-2r)} (-1)^{\sigma_n-2r},
\eeq

The analytic continued integral, valid for the other kinematical
region is given by

\be 
{\cal N}_3^B = \pi^D P_3 \Gamma_3 {\cal F}_3\left(\frac
{2p^+p^-}{p^2n\cdot n^*}\right ) + \pi^D P_4 \Gamma_4 {\cal
F}_4\left(\frac {2p^+p^-}{p^2n\cdot n^*}\right )\label{3som-z}
\ee

where 

\beq
{\cal F}_3\left(\frac{2p^+p^-}{p^2n\cdot n^*}\right)& \equiv &
\sum_{Z_i=0}^\infty \frac{(-\s|Z_{123})(-i|Z_{12}) (j+s+D/2|Z_{2})
(-j-k-D/2|Z_3)}{Z_1!Z_2!Z_3! (l+m+s+D/2|Z_{123}) (l+r+D/2|Z_{123})}
\frac{(D/2+k|Z_1)(D/2+l|Z_{123})}{(j+k+s+D|Z_{12})} \nonumber\\
&\times &\left(\frac{2p^+p^-}{p^2n \cdot n^*} \right)^{ Z_{123}}
\eeq
and

\beq
{\cal F}_4\left(\frac {2p^+p^-}{p^2n\cdot n^*}\right )& \equiv &
\sum_{X_i=0}^\infty \frac{(D/2+k|X_4)(-s|X_{5}) (i+j+k+r+D|X_{456})
(-m|X_{46}) }{X_4!X_5!X_6! (1-l-m-s-D/2|X_{456})} \nonumber\\ &
\times &\frac{(D/2+j+s|X_{6}) (-\s+r|X_{456})} {
(D+j+k+s|X_{46})(1-m+r-s|X_{456})} \left(\frac{2p^+p^-}{p^2n\cdot n^*}
\right)^{ X_{456}},
\eeq

The coefficient factors of external momenta are defined in the table
and the Pochhammer symbols are,
\be 
\G_3 =
(-m-s|-l-D/2)(j+k+s+D|-k-D/2)(-k|k+l+D/2)\frac{(-j|-k-D/2)(-l|k+l+D/2)}{(1+r|l+D/2)}
(-1)^{-l-D/2},
\ee
and

\beq 
\G_4 &=&
(-j-k-D/2|-i)(j+k+s+D|-k-D/2)(\s+D/2|-2\s-D/2+r)(-k|k+l+m+s+D/2)
\nonumber\\ & \times & \frac{(-l|\s-m-s)(-j|j+k+D/2)
}{(1+r|-m-s)}(-1)^{-2i-m-s}.
\eeq

\begin{center}
\begin{tabular}{|c||c|}\hline 
  & Factors \\ \hline\hline 
$ P_1 $ & $(p^2)^{\s-m-r-s}(p^+)^{m+s}(p^-)^r $ \\ \hline 
$ P_2 $ & $\left(\frac{n\cdot n^*}{2} \right)^{-\s+m+r+s}(p^+)^{\s-r} (p^-)^{\s-m-s} $ \\ \hline 
$ P_3 $ & $(p^2)^{D/2+i+j+k}(p^+)^{D/2+l+m+s} (p^-)^{D/2+l+r} \left(\frac{n\cdot n^*}{2} 
\right)^{-l-D/2} $ \\ \hline 
$ P_4 $ & $\left(\frac{n\cdot n^*}{2} \right)^{m+s} (p^+)^{\s-r} (p^-)^{-m+r-s} $ \\ \hline
\end{tabular} 
\end{center}
\begin{center} 
Factors for hypergeometric series representing integral ${\cal N}_3$, eq.(\ref{3som-inv}) and 
eq.(\ref{3som-z}). 
\end{center}

In this section we calculated two-loop \lci{}. \ndim{} is a technique
which can avoid invoking prescriptions, partial fractioning and
integration over components. Moreover, several integrals can be
calculated at once since it is as easy to deal with arbitrary
exponents of propagators as to particular values for them.

\section{Conclusion}

The light-cone gauge is known as the ``trickiest'' of the
non-covariant gauges. This is because unphysical poles, generated by
loop integrals, does appear and may violate causality if care is not
taken as to the correct treatment of those poles. The
cure for such problem came in the form of prescriptions:
Mandelstam-Leibbrandt and causal Cauchy principal value
ones. Recently, we proposed a third way to carry these integrals out,
a method which can abandon prescriptions. \ndim{} is such a
prescriptionless technique and in this work we tested it beyond the
one-loop level. We studied two-loop integrals, which can also be
handled easily, and give results which are compatible with the ones
obtained using ML-prescription.

\acknowledgments{ AGMS gratefully acknowledges FAPESP (Funda\c c\~ao de Amparo \`a 
Pesquisa de S\~ao Paulo) for financial support.}


\begin{thebibliography}{99}


\bibitem{leib-rmp} G.Leibbrandt, Rev.Mod.Phys. {\bf 59} (1987) 1067. 
G.Leibbrandt, {\it Non-covariant gauges: Quantization of Yang-Mills and 
Chern-Simons theory in axial type gauges}, World Scientific (1994).
A.Bassetto, G.Nardelli, R.Soldati,  {\it Yang-Mills  theories in algebraic
non-covariant gauges}, World Scientific (1991). 

\bibitem{mandel}
S.Mandelstam, Nucl.Phys.{\bf B213} (1983) 149. G.Leibbrandt,
Phys.Rev.{\bf D29} (1984) 1699. G.Leibbrandt, J.Williams,
Nucl.Phys.{\bf B566} (2000) 373. G.Leibbrandt, A.Richardson,
C.P.Martin, Nucl.Phys.{\bf B517} (1998) 521. G.Leibbrandt,
Nucl.Phys.{\bf B} (Proc.Suppl.) {\bf 86} (2000) 408.


\bibitem{pimentel}
B.M.Pimentel, A.T.Suzuki, Phys.Rev.{\bf D42} (1990) 2125; Mod.Phys.Lett.{\bf A6} (1991) 2649.

\bibitem{tetrad}
G.Leibbrandt,Phys.Rev.{\bf D30} (1984) 2167. E.T.Newman,R.Penrose,J.Math.Phys.{\bf 3} (1962) 566.

\bibitem{without}
A.T.Suzuki, A.G.M.Schmidt Prog.Theor.Phys. {\bf 103} (2000) 1011; Eur.Phys.J. {\bf C12} (2000) 
361.

\bibitem{leib-cjp}
G.Leibbrandt, Can.J.Phys.{ \bf 64} (1986) 606.

\bibitem{leib-nyeo}
G.Leibbrandt, S-L. Nyeo, J.Math.Phys. {\bf 27} (1986) 627; Z.Phys. {\bf C30} (1986) 501. H.C.Lee, M.S.Milgram, J.Comp.Phys. {\bf 71} (1987) 316.

\bibitem{probing} A.T.Suzuki, A.G.M.Schmidt, R.Bent\'{\i}n, Nucl.Phys.{\bf B537} (1999) 549.


\end{thebibliography}
\end{document}